\documentclass[10pt,letterpaper]{article}
\usepackage[top=0.85in,left=2.75in,footskip=0.75in,marginparwidth=2in]{geometry}

\usepackage[utf8]{inputenc}

\usepackage{cite}
\usepackage{float}
\usepackage{svg}
\usepackage{graphicx}
\usepackage{booktabs}
\usepackage{tabularx}
\usepackage{threeparttable}
\usepackage{siunitx} 

\providecommand{\JournalTitle}[1]{#1}
\usepackage{nameref,hyperref}

\usepackage{microtype}
\DisableLigatures[f]{encoding = *, family = * }

\raggedright
\usepackage{changepage}

\usepackage[aboveskip=1pt,labelfont=bf,labelsep=period,singlelinecheck=off]{caption}

\makeatletter
\renewcommand{\@biblabel}[1]{\quad#1.}
\makeatother

\usepackage{lastpage,fancyhdr,graphicx}
\usepackage{epstopdf}
\fancyheadoffset[L]{2.25in}
\fancyfootoffset[L]{2.25in}

\usepackage{color}

\definecolor{Gray}{gray}{.25}

\usepackage{graphicx}

\usepackage{sidecap}

\usepackage{wrapfig}
\usepackage[pscoord]{eso-pic}
\usepackage[fulladjust]{marginnote}
\reversemarginpar

\begin{document}
\vspace*{0.35in}

\begin{flushleft}
{\Large
\textbf\newline{{Learning from Translation: Seasonal Errors and Feature Importance of the ERA5 Turbulence Predictions}}
}
\newline
\\
Arial Tolentino\textsuperscript{1, 3}, Markus Petters\textsuperscript{2,3}, and Luat T. Vuong\textsuperscript{1, 3,*}
\\
\bigskip
{\bf 1} Department of Mechanical Engineering, University of California Riverside, United States of America 
{\bf 2} Department of Chemical and Environmental Engineering, University of California Riverside, United States of America \\
{\bf 3} Center for Environmental Research and Technology (CE-CERT), University of California Riverside, United States of America
\\

* LuatV@UCR.edu

\end{flushleft}
\section*{Abstract}
Turbulence is a phenomena that is {\it locally} and statistically characterized by measurements, but it is caused by {\it nonlocal}  energy cascades associated with the environment. The presence of turbulence coincides with fluctuations in the refractive index, which impact optical sensing, imaging, and signaling applications. Here, we study the machine learning models that predict near-surface optical turbulence strength $C_n^2$, derived from anemometer-based surface flux measurements through Monin--Obukhov similarity theory, using ERA5 reanalysis data as model inputs. We evaluate the model’s ability to perform temporal extrapolation by training on one year of co-located \(C_n^2\) observations and ERA5 data, and applying the model to ERA5 data from other years at the same site to reconstruct a multi-year time series. We compare the predictions across Southern California and New York. In spite of varying weather and terrain, the ML models show consistent performance and seasonal behavior across training years, revealing trends in both predictive accuracy and ERA5 feature dependence. All models show greater (or reduced) correlation, faster (or slower) convergence, and lower (or higher) prediction errors in the summer (or winter). However, some ERA5 features drive predictions in New York but not California and vice versa, and such feature dependence depends on the season. Seasonal error and feature trends suggest that turbulence is affected by atmospheric composition or other seasonal environmental considerations that are not currently monitored by ERA5. We find, regardless of terrain, the primary feature of importance to turbulence prediction is solar radiation, which underlines the central role of radiative energy transfer in driving atmospheric turbulence. We point toward physics-informed ML translation and feature selection as tools for improving the generalizability of data-driven models.


\section*{Introduction}
\begin{figure}[tbh]
\centering\includegraphics[width= .75\linewidth]{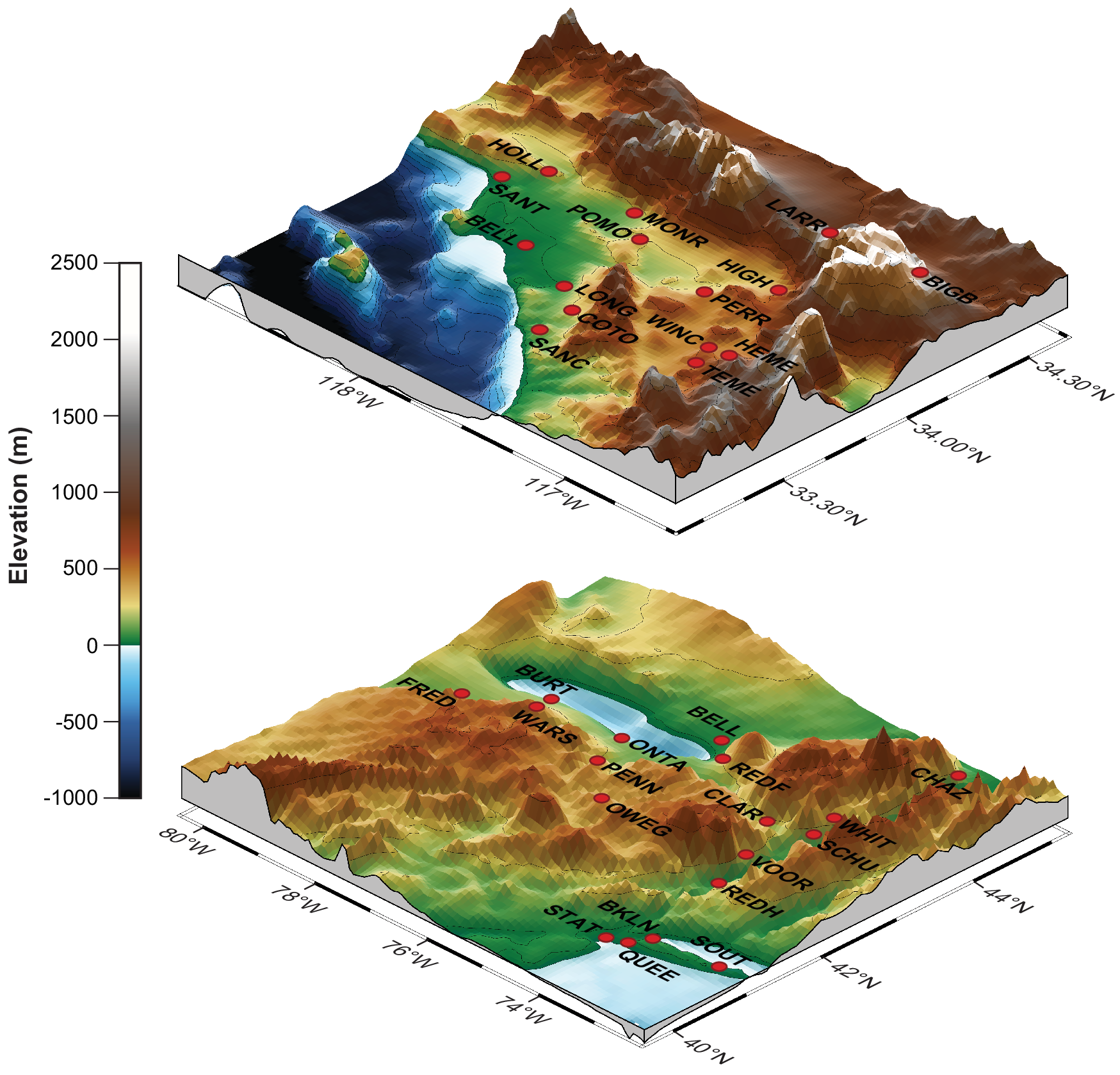}
\caption{3D elevation maps of the two study regions with the anemometer data measurement station locations shown in red. In Southern California (SCA), most of the sites are between the ocean and mountains (left), while in New York State (NYS), the station topography varies and include lakes, valleys, and ocean peninsula (right).}\label{topofig}
\end{figure}

The presence of atmospheric turbulence accompanies fluctuations in the refractive index, which impact optical sensing, imaging, and signaling applications. Atmospheric optical turbulence (OT), quantified by the refractive index structure parameter $C_n^2$, governs the propagation of optical beams through the lower atmosphere. Its magnitude reflects the intensity of refractive index fluctuations within the atmospheric boundary layer. These fluctuations are driven by both thermally generated (buoyancy-driven) and mechanically generated (shear-driven) turbulent mixing within the atmospheric boundary layer, and vary over several orders of magnitude, typically between $10^{-17}$ and $10^{-12}\,\mathrm{m}^{-2/3}$, depending on season, time of day, geographic location, altitude, and local meteorological conditions \cite{Celik2024, Grose2023}. Accurate estimation and forecasting of $C_n^2$ are essential for applications including free-space optical communications, directed-energy systems, adaptive optics, and remote sensing, where turbulence-induced wavefront distortions degrade system performance.

Recent advances in ML have introduced flexible data-driven alternatives capable of learning the nonlinear relationships between meteorological predictors and observed turbulence strength. Neural network models have achieved reliable domain performance for surface-layer $C_n^2$ estimation \cite{Wang2016}, and more advanced architectures have improved generalization to meteorological conditions outside those represented in the training data \cite{Celik2024}. Recurrent neural networks have also been applied to short horizon forecasting using prior meteorological and turbulence observations \cite{Grose2023}. Despite these advances, most studies focus on single-site implementations or short time spans. A central unresolved question is the extent to which learned turbulence meteorology relationships remain stable across multiple years and across distinct climatological regimes.

Prior work establishes reliable approaches for predicting turbulence. In \cite {Wang2016}, an artificial neural-network model predicts near-surface $C_n^2$ more accurately than the classic Wyngaard-derived Monin–Obukhov Similarity Theory (MOST)-based formulation \cite{Wyngaard1971}, even during weak or moderate turbulence conditions. ML models highlight that MOST-based models frequently fail to capture the sharp magnitude and timing of the $C_n^2$ variability during diurnal transition periods. In \cite{Jellen2021} it is shown that polynomial and random-forest models track measured $C_n^2$ under relatively steady atmospheric conditions, but exhibit reduced accuracy during periods of increased variability and diurnal transition. Variations in humidity, along with changes in wind and radiative forcing, are associated with these shifts in model performance. Based in the MCP methodology developed by Wang and Basu \cite{WangBasu2016 } Pierzyna et al., \cite{Pierzyna2025} introduced the OTCliM framework which establishes a measure–correlate–predict methodology linking large scale meteorological drivers from reanalysis products to near-surface OT. 

Building on Wang and Basu, we study the temporal robustness and seasonal structure of the ML-based relationships. Specifically, we examine whether models trained on one calendar year of data can reliably generalize to other years at the same site, and whether this generalization differs across contrasting climatic regions. We apply a round-robin temporal extrapolation framework to two distinct regions of the United States: Southern California (SCA), characterized by coast and mountain terrain with comparatively modest seasonal variability, and New York State (NYS), which exhibits stronger seasonal contrasts and broader inter-annual variability. The terrains and locations of these two domains are shown in Fig. \ref{topofig}.

Motivated by clear seasonal structures in the observed $C_n^2$ distributions, we further decompose annual models into seasonally stratified models to assess regime-dependent predictions and features. This pipeline allows us to test whether turbulence meteorology relationships are temporally stationary or instead reorganize across seasonal boundary layer regimes. We demonstrate that prediction error is governed by seasonal shifts, with higher prediction errors in the winter, regardless of training year. Isolating this regime dependence opens a direct path toward $C_n^2$ models that incorporate seasonally-stratified boundary-layer physics rather than relying on climatologically stationary assumptions.
 
\section{Data and Methodology}

\subsection{Study Regions and $C_n^2$}
Domain 1 corresponds to SCA stations, is characterized by   terrain, strong inland heating, and coastal influences. Domain 2 corresponds to NYS, which spans inland, coastal, rural, and urban environments and experience pronounced seasonal variability driven by synoptic scale weather systems and strong winter/summer contrasts. Both domains include multiple ground-based stations with multi-year records spanning 2018-2022. Regional differences in topography and climatology provide a testbed for assessing inter-annual and seasonal generalization of optical turbulence models.

The SCA station data are obtained from the California Irrigation Management Information System (CIMIS). For NYS, surface flux data are obtained from NYS Mesonet flux stations \cite{Brotzge2020}. CIMIS is a statewide operational network of automated weather stations designed primarily for agricultural and hydrological monitoring, providing hourly observations of surface meteorological variables including air temperature, humidity, wind speed and direction, and solar radiation. The NYS Mesonet is a high-density statewide atmospheric observing network that includes eddy covariance instrumentation intended for weather prediction. Both CIMIS and Mesonet datasets provide the hourly meteorological inputs required to construct the optical turbulence (OT) proxy used in this study. The table summaries of station location, characteristics, and preprocessing variables for both networks are provided in the Supplementary Document.

From the anemometer data, $C_n^2$ is parameterized using Monin–Obukhov similarity theory (MOST), which links turbulent structure parameters to surface fluxes and thermodynamic gradients within the constant flux surface layer \cite{Monin2009BasicLO, Tichkule2012, Muschinski2023, Wyngaard1971, Pierzyna2024}. Because high-frequency (10 Hz) sonic anemometer measurements are not available at all stations, near-surface OT is approximated using MOST under neutral stability assumptions. Sensible heat flux $H$ ($\mathrm{W\,m^{-2}}$), friction velocity $u_*$ ($\mathrm{m\,s^{-1}}$), and air density $\rho$ ($\mathrm{kg\,m^{-3}}$) are used together with air temperature $T$ and relative humidity to estimate the temperature scaling parameter. The Monin--Obukhov length is 
\begin{equation}
L = -\frac{u_*^{3}T}{\kappa\,g\,\overline{w'T'}}
\end{equation}
where $u_*$ is the friction velocity, $\kappa \approx 0.40$ is the von Kármán constant, $T$ is the temperature in K, $g\approx 9.81$ m s$^{-2}$ is the acceleration of gravity, and $\overline{w'T'}$ is the covariance between the vertical velocity fluctuations and temperature fluctuations.  The value of $C_T$ is defined, 
\begin{equation}
C_T^{2}(z)=\frac{T_*^{2}}{z^{2/3}} \Phi_T \left( \frac{z}{L} \right)
\end{equation}
where $z$ is the measurement height, $T_* = \frac{\overline{w'T'}}{u_*}$ is the temperature scale, $\Phi_T \left( \frac{z}{L} \right)$ is the Monin-Obukhov similarity function for temperature \cite{Wyngaard1971}. $C_T$ is used to calculate the changes in the refractive index,  
\begin{equation}
C_n^{2}=
C_n^2 = \left(\frac{\partial n}{\partial T}\right)^2 C_T^2 + 2\frac{\partial n}{\partial T}\frac{\partial n}{\partial Q} C_{TQ} + \left(\frac{\partial n}{\partial Q}\right)^2 C_Q^2,
\end{equation}
where $C_T^2$ ($\mathrm{K}^2\,\mathrm{m}^{-2/3}$) is the temperature structure parameter, $C_Q^2$ ($\mathrm{m}^{-2/3}$) is the humidity structure parameter, 
and $C_{TQ}$ ($\mathrm{K}\,\mathrm{m}^{-2/3}$) is the temperature--humidity covariance structure parameter \cite{wesely1976combined}.

The resulting proxy represents near-surface OT strength at approximately $z = 10\,\mathrm{m}$ above ground level. Several quality control procedures are applied to ensure physically plausible estimates. Periods with weak turbulence, defined here as conditions with low wind speed (\(WS < 1\,\mathrm{m\,s^{-1}}\)) or low friction velocity (\(u_* < 0.10\,\mathrm{m\,s^{-1}}\)), are excluded to satisfy Taylor’s frozen turbulence hypothesis, which assumes that turbulent structures are carried past the sensor by sufficiently strong mean flow. The threshold \(u_* \geq 0.10\,\mathrm{m\,s^{-1}}\) is therefore used to retain well-mixed conditions and remove less well-established turbulence regimes. The use of $u_*$ to distinguish between well-mixed and poorly mixed conditions is standard in eddy covariance smoothed procedures \cite{Papale2006}, and provides quality control over the experimental \(u_*\) data.

In our work, $u_*$ data is captured via anemometers at SCA and NYS stations. Data points flagged as missing, containing no value, indicating a sensor outage, or exceeding sensor thresholds are removed. Finally, we exclude extreme outliers likely associated with sensor artifacts. Following filtering, hourly proxy values are retained and paired with collocated reanalysis features.

\subsection{ERA5}

Meteorological input data are obtained from the ERA5 global reanalysis \cite{Hersbach2020}, which provides hourly atmospheric fields on a regular grid (horizontal resolution $\sim 0.25^\circ$). ERA5 combines numerical weather prediction output with assimilated observations to produce dynamically consistent large-scale atmospheric fields. Data are retrieved for the period 2018–2022 over a domain encompassing all study sites, defined by a bounding box extending beyond the outermost station coordinates. Domain 1 (SCA) is defined by a latitude--longitude bounding box spanning $32.96^{\circ}\mathrm{N}$ to $34.75^{\circ}\mathrm{N}$ and $118.97^{\circ}\mathrm{W}$ to $116.36^{\circ}\mathrm{W}$. Domain 2 (NYS) is defined by $40.10^{\circ}\mathrm{N}$ to $45.40^{\circ}\mathrm{N}$ and $79.87^{\circ}\mathrm{W}$ to $71.97^{\circ}\mathrm{W}$. Each bounding box is constructed to extend $0.5^{\circ}$ beyond the outermost station coordinates in all directions, ensuring full coverage.

For each station, meteorological variables are extracted from the nearest ERA5 grid point and temporally aligned with the hourly $C_n^2$ values. The selected variables represent physical processes governing near-surface turbulence, including boundary-layer structure, surface fluxes and radiation, near-surface temperature, relative humidity, wind speed, and wind direction at multiple levels, and derived variables such as vertical wind shear and temperature gradients. A detailed list of ERA5 variables used in our model, including variable names and units, is provided in the appendix. Derived features follow the OTCliM framework \cite{Pierzyna2025}, with the addition of relative humidity. Although ERA5 does not resolve station-scale heterogeneity, it captures the large scale meteorological forcing that governs boundary layer turbulence.

Although several predictors exhibit partial correlation, tree-based ML models are well-suited to handling redundant inputs. A broad feature space is intentionally retained to allow the model to identify nonlinear interactions relevant to OT variability. After preprocessing, the final dataset consisted of hourly ERA5 features paired with smoothed $C_n^2$ values across multiple stations and years. This unified dataset forms the basis for model training and evaluation.

\subsection{ML model}
\begin{figure}[tbh]
\centering\includegraphics[width= \linewidth]{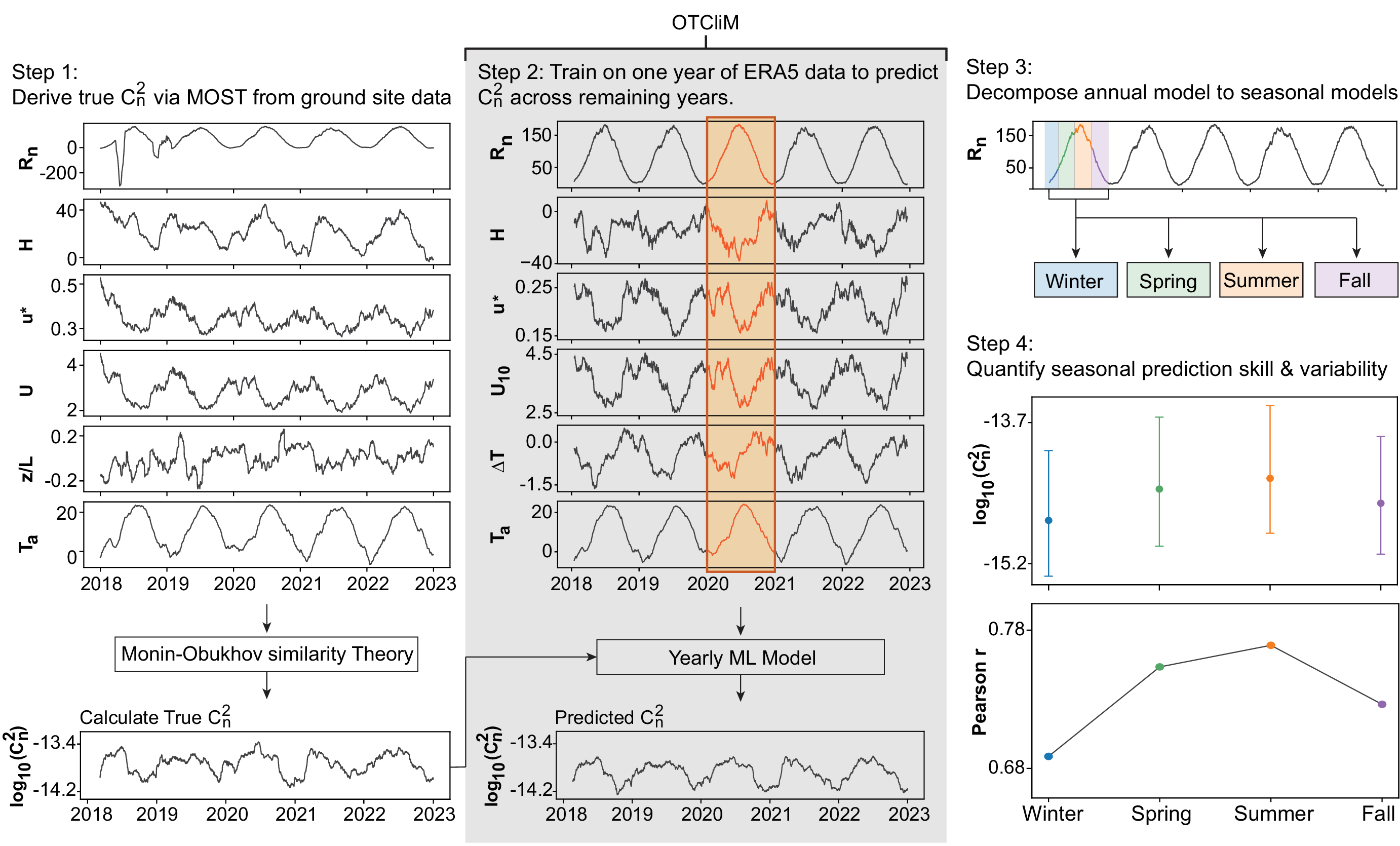}
\caption{Overview of the turbulence modeling workflow. A MOST-derived $C_n^2$ from ground observations is used to train and evaluate a yearly ERA5 based regression model. The annual model is subsequently decomposed into seasonal models to capture regime dependent behavior and extract seasonal features of importance.}\label{ModelOT}
\end{figure}
This study follows the measure–correlate–predict framework summarized in Fig. \ref{ModelOT} (workflow schematic). The methodology consists of four primary steps. First, near-surface $C_n^2$ is derived from ground-based flux measurements using Monin–Obukhov similarity theory (MOST). This step establishes the truth turbulence values used throughout the study. Second, collocated ERA5 reanalysis variables are extracted and paired with the hourly truth values to construct a unified dataset of meteorological predictors and OT response, comprising of 43 input features spanning wind shear, buoyancy, moisture, radiation, and temporal variability. Third, regression models are trained independently for each station using data from a single calendar year. These models are then evaluated on all remaining years using a round-robin temporal extrapolation design. This procedure focuses on the question of temporal generalization, that is whether turbulence and meteorology relationships learned under one year’s boundary layer conditions remain valid under different annual regimes. Finally, motivated by the observed seasonal structure in the $C_n^2$ distributions, the annual modeling framework is decomposed into seasonally stratified models. This allows direct comparison between annual and seasonal training strategies and test regime dependence.

\subsection{Model analyses}
To evaluate temporal robustness, models are trained independently for each station using data from a single calendar year. The target variable is transformed as $\log_{10}(C_n^2)$ and scaled using an inter-quartile range (IQR) normalization fitted exclusively on the training year. For a given station and training year $t$, the trained model is applied to all other years. Predictive performance is quantified using the Pearson correlation coefficient $r$ and the root mean square error (RMSE), computed in scaled $\log_{10}\!\left(C_n^2\right)$ space. This round-robin design allows us to see whether turbulence and meteorology relationships learned under one year’s boundary-layer conditions remain valid under different annual regimes. By repeating this procedure across all stations and years, we quantify inter-annual stability in model performance.

Motivated by pronounced seasonal structure in the observed $C_n^2$ distributions, the annual dataset is partitioned into meteorological seasons: winter (December, January, February), spring (March, April, May), summer (June, July, August), and fall (September, October, November). Independent models are trained and evaluated within each seasonal subset using the same round-robin temporal extrapolation framework. For each station and season, training is performed on one year of seasonal data and tested on the remaining years within that same season. This seasonal stratification isolates boundary layer regimes characterized by distinct stability and forcing conditions. Comparing annual and seasonal model performance enables direct evaluation of whether turbulence meteorology relationships are temporally consistent or change across seasonal transitions.

Finally, we consider the SHapley Additive exPlanations (SHAP), provides additive feature attributions that quantify each input variable's contribution to model predictions \cite{Cremades2025}. SHAP tools reveal the physical importance of input variables in turbulent and thermally driven flows. We use them to understand which environmental variables drive $C_n^2$ predictions.

\section{Results}

\subsection{Seasonal Structure of Observed $C_n^2$}

\begin{figure}[tbh]
\centering\includegraphics[width=4in]{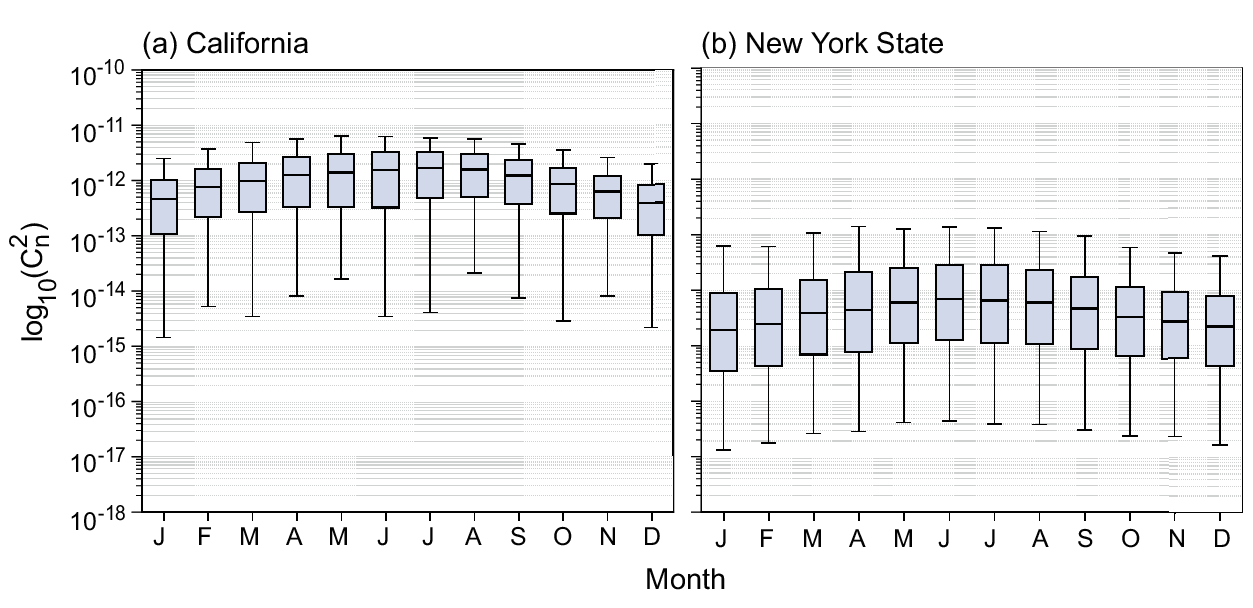}
\caption{ Monthly distributions of $\log_{10}(C_n^2)$ for SCA (a) and NYS (b). Box plots summarize the median, inter-quartile range, and overall variability for each month.} \label{figboxplots}
\end{figure}

To assess baseline seasonal behavior, monthly distributions of observed $C_n^2$ are aggregated across stations within each region. Figure \ref{figboxplots} presents regional box plots summarizing the median, inter-quartile range (IQR; 25th-75th percentiles), and whiskers representing the 5th-95th percentile range of $\log_{10}(C_n^2)$ for SCA and NYS. The use of the 5th-95th percentile range reduces the influence of rare extreme values and emphasizes typical turbulence conditions.

Both regions exhibit a clear annual cycle. Median turbulence levels increase from winter into late spring and summer before declining through autumn. This seasonal variability can be quantified as the difference between winter and summer median $\log_{10}(C_n^2)$, which corresponds to an increase of roughly 0.5 orders of magnitude in both regions. However, variability of this seasonal cycle differ substantially between regions. There is pronounced seasonal contrast in NYS $C_n^2$. Winter months are characterized by lower median values and relatively narrower IQRs compared to other seasons, with median $\log_{10}(C_n^2)$ near -14.6 in winter increasing to approximately -14.1 in summer. This corresponds to a seasonal amplitude of .46 in $\log_{10}$ space. These conditions are consistent with suppressed boundary layer turbulence under stable stratification and reduced convective forcing under cold conditions. In contrast, spring and summer months show elevated medians and broader IQRs, indicating stronger convective forcing and greater variability. The IQR increases from approximately 1.33 in winter to 1.36 in summer, indicating modest but consistent increases in variability associated with convective conditions. 

While SCA also exhibits enhanced $C_n^2$ during the late spring and summer, the seasonal variation is comparatively moderate. Median values remain elevated throughout the year relative to NYS, and winter suppression is less pronounced. Median $\log_{10}(C_n^2)$ ranges from approximately -12.3 in winter to -11.8 in summer, corresponding to a seasonal amplitude of .49 in $\log_{10}$ space, similar in magnitude to NYS. However, across all seasons, SCA exhibits consistently higher median values by roughly two orders of magnitude compared to NYS. Despite having a seasonal amplitude comparable to NYS, the relatively smaller changes in IQR and consistently higher baseline suggest that turbulence in SCA is more persistently maintained throughout the year rather than being strongly driven by seasonal shifts. This is consistent with the region's high annual mean solar irradiance, low atmospheric moisture. In arid inland environments, the surface energy balance 
preferentially partitions incoming radiation into sensible heat flux rather than evapotranspiration, sustaining strong near-surface temperature gradients and consequently elevated refractive index fluctuations throughout the year. This physical mechanism is corroborated by the SHAP feature importance analysis (Section~3.4), which identifies solar radiation as the dominant predictor of 
$C_n^2$ across both regions.

\subsection{Annual Model Temporal Extrapolation Performance}

\begin{figure}[t]
\centering\includegraphics[width= \linewidth]{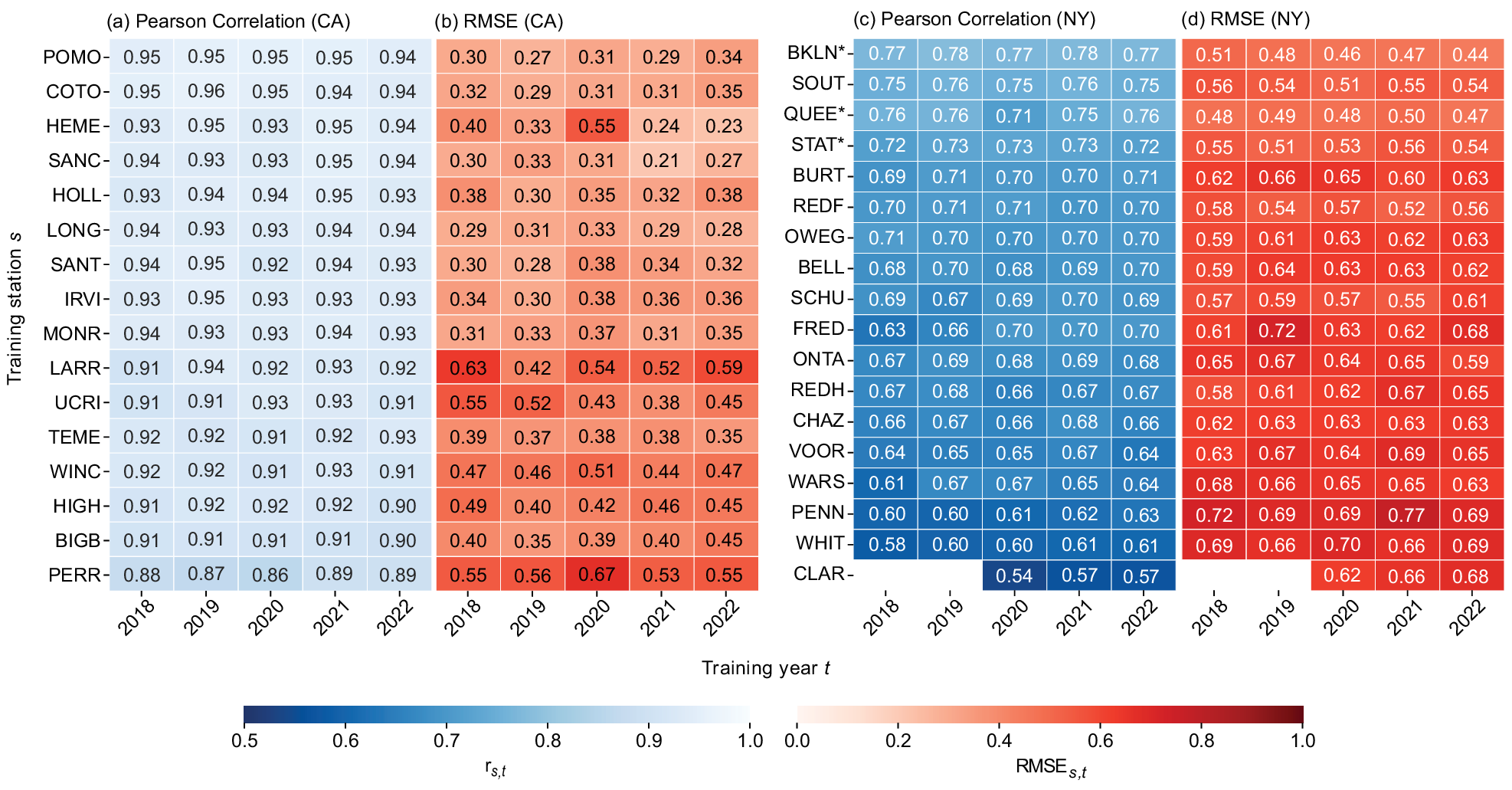}
\caption{Heatmaps of annual station-specific model performance. Pearson correlation and RMSE for (a-b) SCA and (c-d) NYS. which are trained on a single year and evaluated on all remaining years at the same station. Rows indicate training stations and columns indicate the calendar year used for model training. The annual models trained on SCA compared to those trained with NYS data exhibit higher correlations and lower RMSE values.}
\label{heatmaps}
\end{figure}

Figure \ref{heatmaps} presents annual model performance using the temporal extrapolation framework described in Section 2. For each region, rows correspond to training stations and columns correspond to the calendar year used for model training. Each cell represents performance evaluated on all remaining years at the same station. Pearson correlation coefficient $r$ and root mean square error (RMSE) are shown in separate panels. All metrics are computed using $\log_{10}(C_n^2)$ and subsequently scaled using an IQR normalization (Section~2).

A clear regional contrast in model performance is evident. The SCA station models exhibit consistently high correlation values across training years, generally exceeding $r \approx 0.9$. Across all station-year combinations, SCA models achieves a mean correlation of $0.93 \pm 0.02$, with values ranging from 0.86--0.96. Performance variability across training years is minimal, and RMSE values remain comparatively low. The mean SCA RMSE is $0.39 \pm 0.10$, with values ranging from 0.21--0.67. In contrast, NYS station models display lower correlations and greater inter-annual variability. Across all station-year combinations, NYS has a mean correlation of $0.68 \pm 0.05$, with values ranging from 0.54--0.78 while mean RMSE is $0.61 \pm 0.07$, ranging from 0.44--0.07. Correlation values typically range from approximately $r \sim$ 0.6--0.8 with noticeable dependence on the selected training year. RMSE values are correspondingly higher and exhibit stronger variability across training–testing combinations. Several stations show increased sensitivity to specific training years, suggesting that the data captured on boundary layer regimes in NYS are less interchangeable year-to-year.

These cross-year generalization trends are consistent with the seasonal model performance observed at individual stations. Although both regions showed comparable seasonal amplitude in the observed $C_n^2$ distributions, NYS exhibited substantially broader IQRs and greater intermittency, which likely weakens the transferability of year-specific feature-target relationships. In particular, models trained under one seasonal regime may fail to capture the distinct forcing conditions of another, especially during transitions between stable and convective boundary-layer states. This interpretation is supported by the larger mean spread in correlation across stations for NYS ($\sim 0.053$) compared to SCA ($\sim 0.019$) indicating that model skill is less consistent from site to site in NYS. In contrast, despite exhibiting a seasonal amplitude comparable to NYS, SCA shows reduced variability, which contributes to stronger generalization across years. The combination of persistently elevated turbulence and reduced distributional spread in SCA appears to produce more stable annual relationships, allowing models trained on one year to extrapolate more successfully to others.

\subsection{Seasonal Model Performance} 

\begin{figure}[bth]
\centering\includegraphics[width= .85\linewidth]{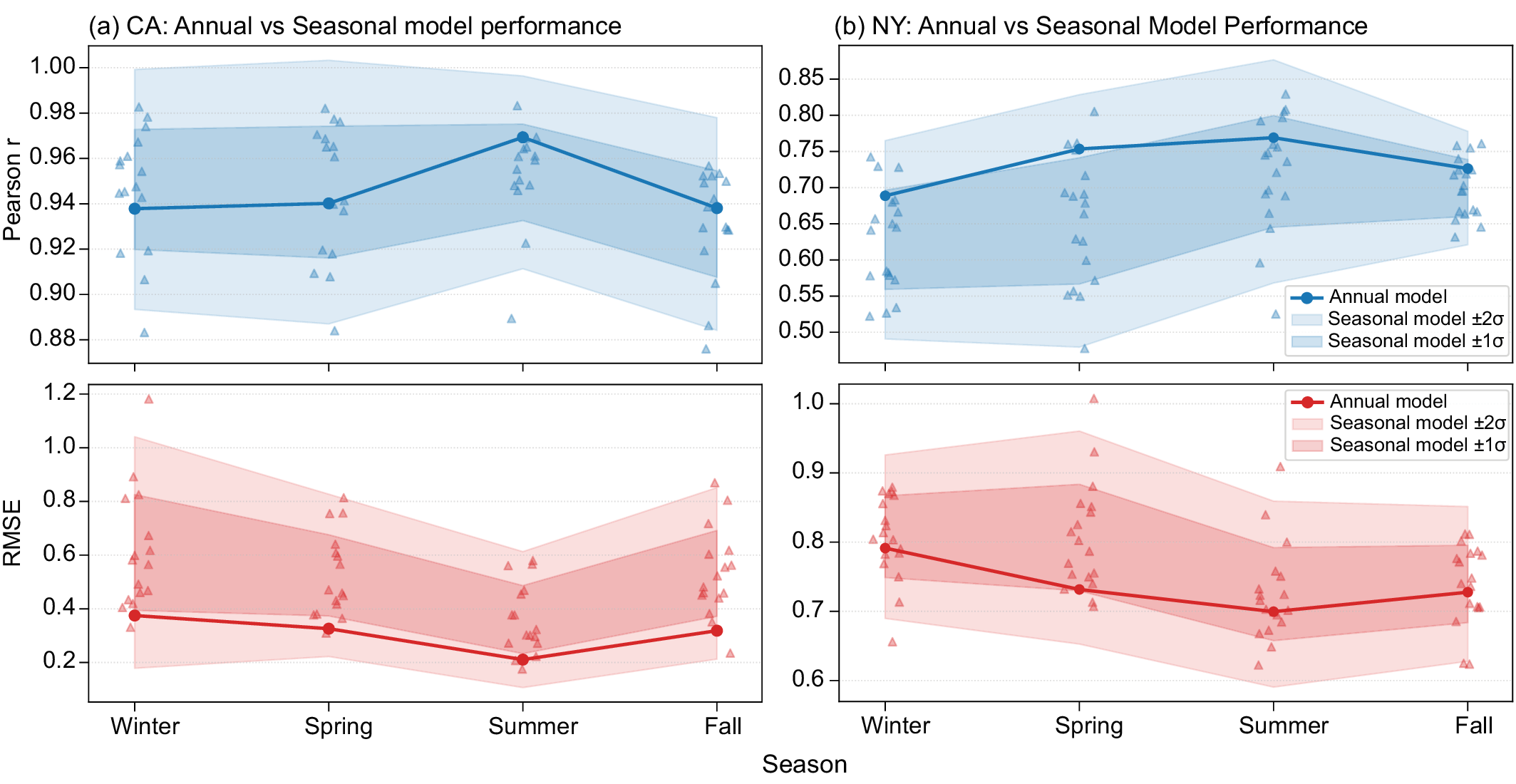}
\caption{Seasonal model performance for SCA (left) and NYS (right). Pearson correlation (top panels) and RMSE (bottom panels) for annually trained models (solid lines) and seasonally trained models (shaded $\pm 1\sigma$ and $\pm 2\sigma$ bands). Points denote station MCP performance.}\label{Seasonalfig}
\end{figure}

To evaluate regime dependence, annual model performance is compared against seasonally trained models. Figure \ref{Seasonalfig} summarizes Pearson correlation $r$ (top panels) and RMSE (bottom panels) for SCA (left) and NYS (right). Solid lines represent the annually trained model, while shaded bands indicate $\pm 1\sigma$ and $\pm 2\sigma$ variability across stations for the seasonally trained models. Individual points denote station level mean cross-prediction (MCP) performance.

\begin{figure}[h!]
\centering\includegraphics[width= .8\linewidth]{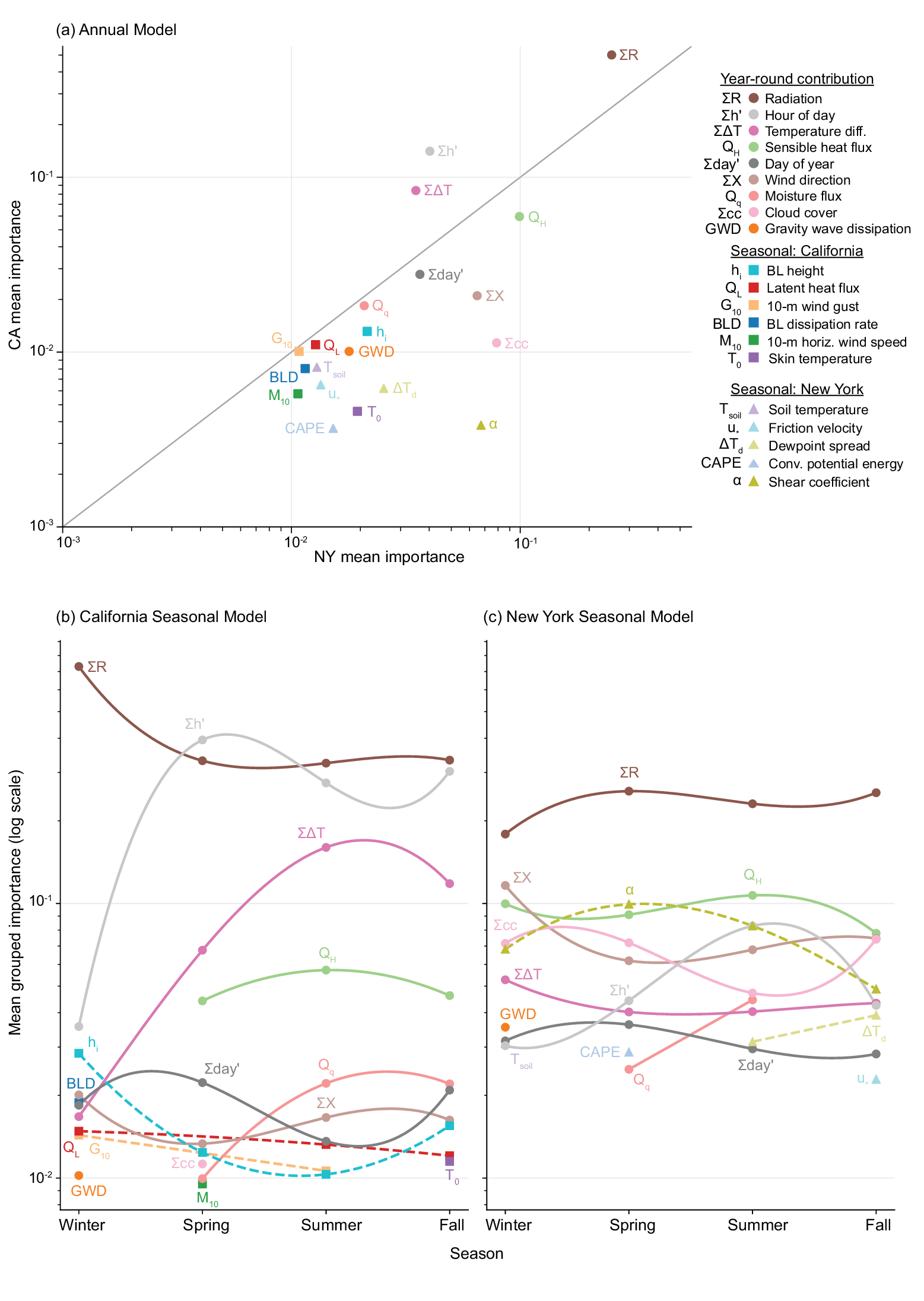}
\caption{(a) Annual feature importance comparison between SCA and NYS. 
Seasonal evolution of grouped feature importance for (b) SCA and (c) NYS, respectively. }\label{SHAPfig}
\end{figure}

In SCA, correlations remain consistently high across all seasons, ranging from $r \approx 0.88$ in winter to $r \approx 0.98$ in summer, with a seasonal spread of $\Delta r \approx 0.02$--$0.04$ across stations. RMSE values 
similarly show limited seasonal variation, remaining within $0.3$--$0.6$ in normalized $\log_{10}(C_n^2)$ units, indicating that turbulence--meteorology relationships are stable throughout the year and that season-specific modeling provides minimal additional benefit over the annually-trained model. RMSE values show similarly limited seasonal variation, indicating that turbulence meteorology relationships are relatively stable throughout the year and that season specific modeling provides minimal additional benefit. By contrast, NYS exhibits pronounced seasonal structure, with correlations lowest in winter and increasing through spring and summer, peaking during the convective season. RMSE follows the opposite pattern, with the largest errors in winter and the smallest in summer. 

\subsection{Feature Importance}
Figure \ref{SHAPfig} shows the normalized SHAP feature importance. Panel (a) shows the relative overall importance between SCA and NYS, while panels (b) and (c) show the top ten SHAP features as they vary with importance across domain and seasons, highlighting temporal shifts. SHAP values quantify the contribution of each feature to the model prediction relative to a baseline prediction. Feature importance refers to the mean absolute SHAP value. These SHAP values are then normalized such that the sum of the SHAP values for an individual model equals one, so that each value represents the fractional contribution of a feature to the total model importance. These unit-normalized SHAP values are plot on a logarithmic scale to emphasize differences across orders of magnitude. 

In Fig. \ref{SHAPfig}(a), each point represents the mean importance of a feature in NYS ($x$-axis) and SCA ($y$-axis). Points along the 1:1 diagonal indicate equal importance in both regions. Three points above the diagonal correspond to features that are more important in SCA, six points below the diagonal are features that are more important in NYS. Figure \ref{SHAPfig}(b-c) shows the feature importance associated with seasonal models in SCA and NYS. 

Varying feature importance is associated with the seasonal model performance [Fig. \ref{Seasonalfig}], which reveals an additional layer of spatial variability across stations. The full SHAP ranking is provided in the Supplemental Document. Here we summarize the key contrasting trends between domains. As shown in Fig.~\ref{SHAPfig}(a), features shared as top predictors across both regions include radiation, hour of day, temperature difference, sensible heat flux, day of year, wind direction, moisture flux, cloud cover, and gravity wave dissipation. This indicates a common set of boundary-layer drivers that govern $C_n^2$ regardless of climatological regime.

There exists features that are uniquely important in SCA such as boundary layer height, latent heat flux, 10-m wind gust, boundary-layer dissipation rate, 10-m horizontal wind speed, and skin temperature. While solar radiation carries the highest SHAP value in almost all seasonal models, it is the dominant SCA winter feature. The trend of solar radiation importance with boundary layer height and temperature difference is inverted during the spring and summer months. This shows the dominance of surface energy balance and thermally-driven circulation in the arid inland environment. By contrast, features uniquely important in NYS are soil temperature, friction velocity, dew point spread, convective available potential energy, and the shear coefficient which points to the stronger role of surface moisture and mechanical turbulence production in the coastal continental climate. Notably, the distribution of features and their apparent role varies temporally. In SCA, a few features dominate in importance with SHAP values < 0.02, whereas the NYS SHAP values are clustered and range between 0.02 and 0.2. The results suggest that turbulence SHAP feature relationships in NYS are more complex compared to SCA, particularly in their seasonal transitions.

\section{Discussion and Conclusion}

As ML emerges as a tool to find patterns and make predictions, important questions emerge in regards to how much we can rely on the ML model prediction data, i.e., how explainable are the patterns mapped by ML models? Phenomena such as turbulence--where ground truth local measurements are used to infer nonlocal phenomena--are the epitome of such challenges. In our work, we show that prediction error and variability in feature importance are important elements of ML model analyses. In our pipeline, we translate and compare feature dependence over different temporal ranges. This research pipeline is a good first step towards physics-informed ML models, since seasonal changes in the prediction error and feature dependence may point to coupled parameterization in MOST or the presence of hidden variables (or absence of input variables). 

While MOST based formulations are physically grounded and widely used, their validity is restricted to the surface layer, whose depth and stability vary diurnally and seasonally \cite{Sadot1992, Wang2015}.  Alternatives to MOST exist but have not been explored here \cite{Muschinski2016, Casasanta2025}. Similarity functions in MOST diverge strongly in stable, transitional, and heterogeneous conditions, leading to large uncertainty in flux and turbulence predictions \cite{Li2011}, which may explain why SCA models exhibit both higher $C_n^2$ and lower prediction errors as associated with Pearson correlation and RMSE. Standard macro-meteorological formulations systematically misrepresent turbulence strength in a near-maritime environment \cite{Jellen2020} which may relate to high prediction variations in prediction for Hemet (HEME), Lake Arrowhead (LARR), Perris (PERR) in Southern Caliornia, and Burt (BURT), Penn Yan (PENN) and Fredonia (FRED) in NY. This said, using spectral filtering and error-weighted curve fitting across multiple field datasets, MOST similarity functions hold across a wide stability range \cite{Kooijmans2016}. On the other hand, turbulence statistics are also well-known to be anisotropic \cite{Korotkova2021, Adams2025a, Kulikov2025}, and MOST stems from the ideal assumption of isotropy \cite{Tatsumi1980}. Therefore, seasonal variations in the anisotropic shifts in turbulence may also relate to seasonal prediction errors.

Maritime environments may also influence seasonal prediction accuracies and feature dependence via humidity through several factors, including hardware measurements, noise, or sensor drift \cite{Ldi2005, Odhiambo2009, ElMadany2013, Loescher2005}. Humidity also introduces a memory-like effect \cite{Alduchov1996}, which may related to  the urban boundary layer \cite{Barlow2014, Oke2017}. Urban flux measurements are not necessarily representative of the wider turbulence environment, since source areas shift with stability, height, and wind direction. As such, the atmospheric composition related to hydroscopic aerosols \cite{Ravichandran2026}, may also carry a role in relation to the results here. It has been shown that humidity has its own turbulent structure, with small fluctuations in the mixed layer but large, skewed, and intermittent fluctuations near the entrainment zone—especially when moist or dry layers mix \cite{Muppa2015}. This might help explain why our $C_n^2$ models are less reliable in very humid, stable, or transition periods.

In conclusion, we establish common ML-learned patterns in the turbulence prediction and environment and variability across stations in NYS and SCA. While radiation, time/date, wind direction, heat and moisture flux, and cloud cover are the primary feature groups of high importance across all stations, others emerge as dominant only at specific locations or seasons, reflecting differences in local surface characteristics, meteorological regimes, or measurement environments. Our results highlight the conceptual approach using ML as a tool to understand measurements and models rather than a tool for prediction alone.

\section{Acknowledgment}
A.T. and M.P. are funded by a Winston Chung Global Energy Center Seed grant. L.T.V. gratefully acknowledges helpful discussions with Sunilkumar K and additional support for A.T. through AFSOR FA9550-24-1-0027 DEF. 

\section{Data availability} Data underlying the results presented in this paper are available in Ref. \cite{Pierzyna2025}.

\section{Disclosures} The authors declare no conflicts of interest.
\\


\end{document}